\documentclass[5p]{elsarticle}

\usepackage{lineno,hyperref}
\usepackage{graphicx}% Include figure files
\usepackage[english]{babel}
\usepackage{units}
\usepackage{color}
\usepackage{epstopdf}
\usepackage{amsmath}
\usepackage{color}

\biboptions{sort&compress}

\begin{document}

\begin{frontmatter}

\title{Spin and charge transport in graphene-based spin transport devices with Co/MgO spin injection and spin detection electrodes}

\author{F. Volmer, M. Dr\"{o}geler, G. G\"untherodt, C. Stampfer, B. Beschoten}
\address{2nd Institute of Physics and JARA-FIT, RWTH Aachen University, D-52074 Aachen, Germany}
\ead{bernd.beschoten@physik.rwth-aachen.de}

\begin{abstract}
In this review we discuss spin and charge transport properties in graphene-based single-layer and few-layer spin-valve devices. We give an overview of challenges and recent advances in the field of device fabrication and discuss two of our fabrication methods in more detail which result in distinctly different device performances. In the first class of devices, Co/MgO electrodes are directly deposited onto graphene which results in rough MgO-to-Co interfaces and favor the formation of conducting pinholes throughout the MgO layer. We show that the contact resistance area product ($R_cA$) is a benchmark for spin transport properties as it scales with the measured spin lifetime in these devices indicating that contact-induced spin dephasing is the bottleneck for spin transport even in devices with large $R_cA$ values. In a second class of devices, Co/MgO electrodes are first patterned onto a silicon substrate. Subsequently, a graphene-hBN heterostructure is directly transferred onto these prepatterned electrodes which provides improved interface properties. This is seen by a strong enhancement of both charge and spin transport properties yielding charge carrier mobilities exceeding $\unit[20000]{cm^2/(Vs)}$ and spin lifetimes up to 3.7~ns at room temperature. We discuss several shortcomings in the determination of both quantities which complicates the analysis of both extrinsic and intrinsic spin scattering mechanisms. Furthermore, we show that contacts can be the origin of a second charge neutrality point in gate dependent resistance measurements which is influenced by the quantum capacitance of the underlying graphene layer.
\end{abstract}


\end{frontmatter}

\section{Introduction}

In spin-based electronics, three main aspects have to be considered when exploring suitable spin transport materials and material combinations: (1) electrical injection and detection of spins, (2) their manipulation, and (3) the transport of spins in the material  \cite{Wolf16112001,RevModPhys.76.323,ISI000249789600001}. What makes graphene a promising material in the field of spintronics, is its unique spin transport performance in particular at room temperature \cite{GrapheneSpintronics} where spin lifetimes of up to $\unit[3.7]{ns}$ \cite{Droegeler2014} and spin diffusion length of $\unit[12]{\mu m}$ \cite{PhysRevLett.113.086602} have been measured by means of electrical Hanle spin precession measurements in non-local spin-valve devices. The corresponding charge carrier mobilities in these devices are above $\unit[20000]{cm^2/(Vs)}$ \cite{Droegeler2014,PhysRevLett.113.086602}. Other interesting materials in the field of spintronics, e.g. Si, also exhibit nanosecond spin lifetimes at room temperature but fail short to graphene in respect to the spin diffusion lengths and charge carrier mobilities \cite{SiliconSpintronics}. We note that we only compare device and material properties from electrical spin precession measurements at room temperature. Less invasive spin sensitive methods such as electron spin resonance or optical pump-probe methods can yield much longer spin lifetimes, especially at low temperatures \cite{SiliconSpintronics,GrapheneSpintronics,Wu201061,PSSB:PSSB201350201}.

{Graphene has an extraordinary band structure \cite{RevModPhys.81.109} and a weak intrinsic spin-orbit coupling at energies close to the Dirac point \cite{Pesin2012}. Often, this property is mentioned as a favorable aspect for graphene spintronics, because for the most prominent spin relaxation mechanisms the spin relaxation rate scales with the spin-orbit coupling strength \cite{RevModPhys.76.323,ISI000249789600001}. Accordingly, initial calculations promised quite long spin lifetimes in pristine graphene flakes up to the ms regime \cite{PhysRevB.80.041405}. But the experimental values are orders of magnitude smaller than these predictions and only exhibit spin lifetimes in the range of $\unit[20]{ps}$ to $\unit[3.7]{ns}$ at room temperature \cite{Tombros2007,PhysRevLett.113.086602,PhysRevB.80.214427,doi:10.1021/nl301050a,doi:10.1021/nl2042497,PhysRevB.86.161416,PhysRevB.89.035417,han222109,PhysRevLett.105.167202,PhysRevLett.104.187201,doi:10.1021/nl301567n,PhysRevB.87.075455,Avsar2011,PhysRevLett.107.047206,PhysRevB.88.161405,Droegeler2014,PhysRevB.90.165403,PhysRevB.84.075453,abel:03D115,PhysRevB.87.081405,Neumann2013,APEX.6.073001,Kamalakar2014,Kamalakar2015,1.4893578,FU_JAP_2014,1882-0786-7-8-085101,Yamaguchi2012849,doi:10.1021/acsnano.5b02795}. More elaborated calculations which included novel spin scattering mechanisms such as resonant scattering by magnetic impurities \cite{PhysRevLett.112.116602} or entanglement between spin and pseudospin by random spin orbit coupling \cite{Tuan2014} can explain these short spin lifetimes. Nevertheless, final answers about both the limiting spin relaxation mechanism and the maximal achievable spin lifetime in graphene are still missing. Considering the latter, it is interesting to mention that electron spin resonance (ESR) experiments in synthesized graphene flakes yield spin lifetimes of conduction electrons in the range of $\unit[30-65]{ns}$, despite of a significant defect density in the studied samples \cite{AugustyniakJablokow2013118,ESR2014}. It has been suggested that the much longer spin lifetimes in ESR is due to the fact that the graphene sheets are free from substrate effects and metallic electrodes \cite{doi:10.1021/nn302745x}.}

{There is a large number of publication from different groups about electrical injection and detection of spins in graphene (e.g. \cite{Tombros2007,PhysRevLett.113.086602,PhysRevB.80.214427,doi:10.1021/nl301050a,doi:10.1021/nl2042497,PhysRevB.86.161416,PhysRevB.89.035417,han222109,PhysRevLett.105.167202,PhysRevLett.104.187201,doi:10.1021/nl301567n,PhysRevB.87.075455,Avsar2011,PhysRevLett.107.047206,PhysRevB.88.161405,Droegeler2014,PhysRevB.90.165403,PhysRevB.84.075453,abel:03D115,PhysRevB.87.081405,Neumann2013,APEX.6.073001,Kamalakar2014,1.4893578,FU_JAP_2014,1882-0786-7-8-085101,Yamaguchi2012849,doi:10.1021/acsnano.5b02795,Kamalakar2015}). But although electric fields from back and/or top gates or biases along the graphene channel can strongly modify spin and charge transport properties, the actual spin precession in these experiments is always triggered by external magnetic fields. The reason for this is that the before-mentioned weak spin-orbit coupling in graphene is also a mixed blessing because for spin manipulation, e.g. via the Bychkov-Rashba effect, a strong spin-orbit coupling is needed \cite{RevModPhys.76.323,ISI000249789600001}, which, on the other hand favors spin dephasing and spin scattering as it has been explored in III-V semiconductors \cite{Kuhlen_2012,Stepanov_2014}.} Therefore, one emerging topic in graphene-based spintronic research is the partial functionalization of graphene with the goal to achieve other ways for spin manipulation, e.g. by electrostatic gating. The aim is to use high quality graphene parts as leads for efficient spin transport whereas the spin manipulation is realized in a functionalized part of the graphene device with enhanced spin-orbit coupling. Routes to increase the spin-orbit coupling include the spin-orbit proximity effect, in which additional materials such as the two dimensional transition metal dichalcogenides (e.g. tungsten disulfide \cite{Avsar2014}) or ferromagnetic insulators (e.g. EuO \cite{doi:10.1021/nn303771f}) are put in direct contact to graphene. Also hydrogenation of graphene \cite{Balakrishnan2014} or the deposition of heavy adatoms \cite{Marchenko2012} can yield enhanced spin-orbit coupling in graphene, but as a drawback, these approaches are already known to strongly change its band structure \cite{PhysRevB.83.165433, Marchenko2012}.

Very recently, an excellent review article about graphene spintronics compiled a significant part of both theoretical and experimental work which has been carried out on spin phenomena in graphene \cite{GrapheneSpintronics}. In particular, this review focuses on how to measure spin transport in graphene, the spin-orbit coupling in pristine and modified graphene, magnetic moments from defects and adatoms, and the open question about which spin relaxation mechanisms limit spin transport in graphene. In our article, we expand the review to open questions and challenges in the experimental field of work. In section \ref{NewMethods} we discuss new routes in device fabrication by the usage of novel transfer techniques and outline the issue of device contaminations during fabrication. Then, we cover the influence of metallic contacts on the underlying graphene in section \ref{ImpactOnGraphen} and present new results on the appearance of a contact-induced second charge neutrality point in gate dependent resistance measurements. Next, the reliability of extracted values for both spin lifetimes and charge carrier mobilities is discussed in sections \ref{lifetime} and \ref{Mobilities}, respectively. In section \ref{SpinRelaxationMechanism}, we briefly comment on {some experimental studies which were used to investigate the relevant spin relaxation mechanism in graphene.}

%%%%%%%%%%%%%%
\section{Methods of device fabrication}
\label{NewMethods}

\begin{figure*}[tb]
\centering
	\includegraphics{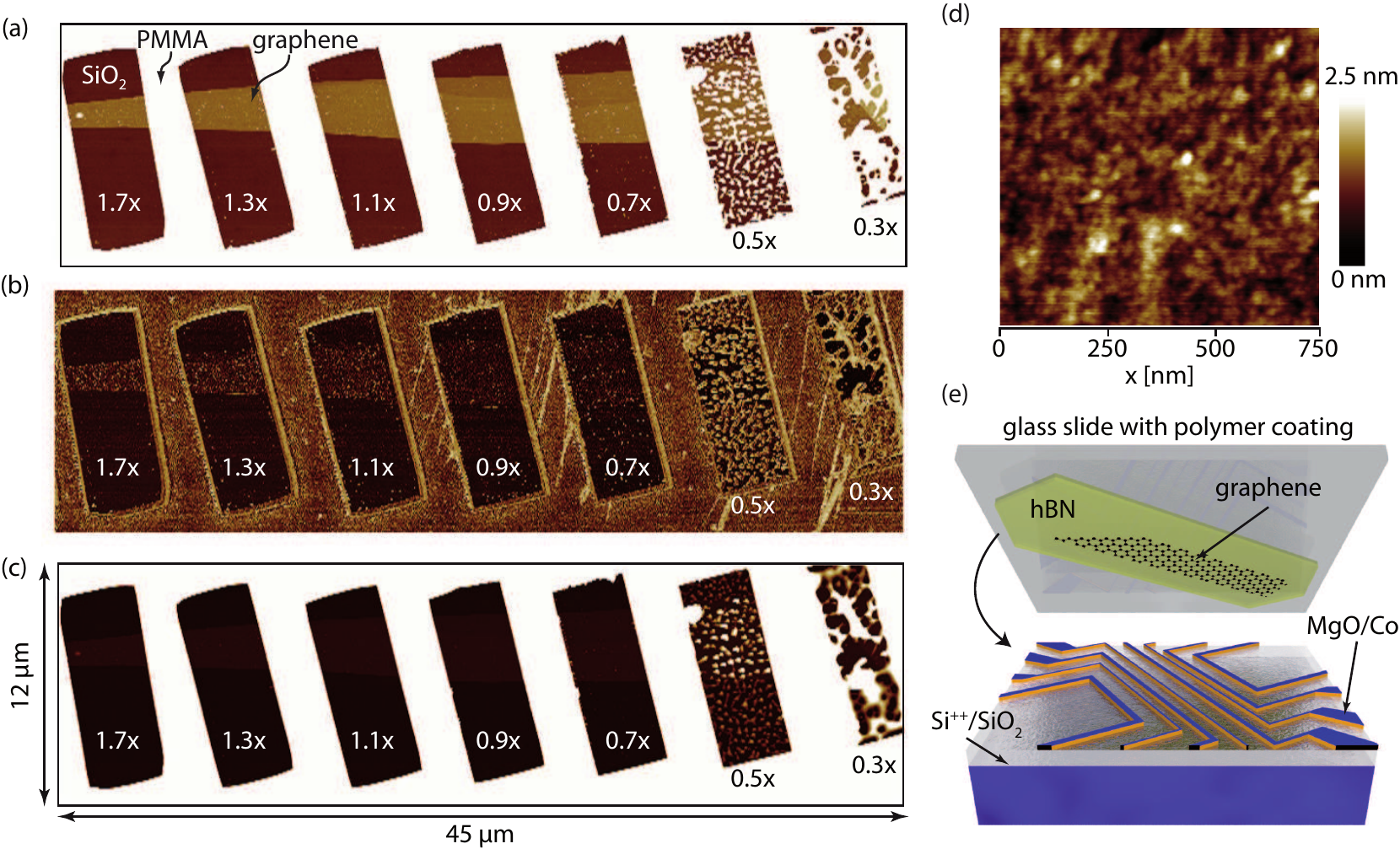}
	\caption{(Color online) (a) AFM image of a graphene flake with spin coated PMMA resist. The fields were written by e-beam lithography with different dosages as given by the denoted factors times the reference dosage of $\unit[100]{\mu C / cm^2}$. The scale is adjusted to maximize the contrast of the graphene flake. (b) In the phase signal of the AFM the resist residues on top of graphene are clearly visible. (c) Image from (a), but now the scale is optimizes to visualize the underdeveloped resist residues. (d) AFM image of a 3~nm thick MgO layer on top of graphene demonstrating the Volmer-Weber island growth: rms roughness of 0.4~nm and peak-to-peak values of 2~nm. (e) Transfer technique for the fabrication of graphene spin-valves from the bottom-up approach in ref.~\cite{Droegeler2014}. Figures reproduced with permission from: (d), ref. \cite{PhysRevB.90.165403}, \copyright 2014 American Physical Society; (e), ref. \cite{Droegeler2014}, \copyright 2014 American Chemical Society.}
	\label{figure1}
\end{figure*}

 {In recent years, it has been demonstrated that electrical injection and detection of spins in graphene can be accomplished by a variety of different electrode materials, such as Al$_2$O$_3$/Co \cite{Tombros2007,PhysRevB.80.214427,doi:10.1021/nl301050a,doi:10.1021/nl2042497,PhysRevB.89.035417,PhysRevB.87.081405}, MgO/Co \cite{han222109,PhysRevLett.104.187201,PhysRevLett.107.047206,Avsar2011,PhysRevB.88.161405,PhysRevB.90.165403,Droegeler2014}, submonolayer TiO$_2$/MgO/Co \cite{PhysRevLett.105.167202,doi:10.1021/nl301567n,PhysRevB.87.075455}, TiO$_2$/Co \cite{PhysRevB.86.161416,PhysRevLett.113.086602}, Cu/NiFe \cite{10.1063/1.4733729}, amorphous carbon/Co \cite{10.1063/1.4820586}, fluorinated graphene/NiFe \cite{Friedman2014}, hydrogenated graphene/NiFe \cite{doi:10.1021/acsnano.5b02795}, PTCA/ALD Al$_2$O$_3$/NiFe \cite{Yamaguchi2012849}, only Co \cite{PhysRevB.84.075453,abel:03D115,Neumann2013}, h-BN/NiFe \cite{APEX.6.073001}, h-BN/Co \cite{Kamalakar2014,1.4893578,FU_JAP_2014}, and YO/Co \cite{1882-0786-7-8-085101}. However, as mentioned in the introductory part, the measured spin lifetimes are only in the range of $\unit[20]{ps}$ to $\unit[3.7]{ns}$ at room temperature. This is well below the lifetimes of $\unit[30-65]{ns}$ measured by ESR experiments in graphene flakes without any contacts \cite{AugustyniakJablokow2013118,ESR2014}. Next to these ESR experiments there are also other studies which indicates that an insufficient barrier quality can be the bottleneck for the overall spin transport \cite{PhysRevLett.105.167202,PhysRevB.86.235408,PhysRevB.88.161405,PhysRevB.90.165403}. In the following, we therefore address the challenges and summarize the progress in the field of device fabrication.}

 {First, we discuss several shortcomings when growing electrode material directly onto the graphene surface.} Because often the starting point of a graphene-based spin transport device is a randomly exfoliated graphene flake which is typically deposited onto Si$^{++}$/SiO$_2$. The next fabrication step is usually a lithography process on top of graphene which is needed for the deposition of ferromagnetic electrodes. But as soon as graphene gets into contact with an organic resist, it is extremely difficult to completely remove the contamination of hydrocarbons. This is illustrated in the atomic force microscopy (AFM) images of figures \ref{figure1}(a)-(c), which show one of our earlier but not optimized dosage tests. As a resist, we used PMMA (950K) which was dissolved in ethyl lactate and n-butyl acetate with a thickness of 250~nm after spin-coating and baking. For developing we used isopropyl alcohol and methyl isobutyl ketone with a developing time of 105~s. E-beam writing was performed with an acceleration voltage of 10~kV. The dosages in figures \ref{figure1}(a)-(c) are given in fractions of $\unit[100]{\mu C / cm^2}$. The scale in figure \ref{figure1}(a) was chosen to depicted the flake in high contrast. The resist is underdeveloped for a dosage below $\unit[70]{\mu C / cm^2}$. But also for dosages larger than $\unit[110]{\mu C / cm^2}$ an increasing contamination of the graphene flake can be observed, which can be easier be visualized in the phase signal of the AFM image in tipping mode (figure \ref{figure1}(b)). The difference in contrast can be understood by different attenuations of the AFM cantilever for SiO$_2$, graphene on SiO$_2$, and PMMA on graphene. The increasing contamination for higher dosages can be explained by crosslinking of PMMA \cite{0268-1242-11-8-021} or e-beam-induced defects in the graphene flake \cite{teweldebrhan:013101, 10.1063/1.3502610} which enables a stronger binding with hydrocarbons. We observe the cleanest graphene surface at a dosage of $\unit[90]{\mu C / cm^2}$ and were able to reduce the amount of leftover resist residues even further by increasing the developing time to 210~s.

But even for optimized lithography conditions the graphene flake will be contaminated with hydrocarbons on an atomic scale. Such contaminations are clearly seen by transmission electron microscopy and annealing temperatures of up to $\unit[700]{^{\circ} C} $ are needed to remove the hydrocarbons \cite{doi:10.1021/nl203733r, doi:10.1021/jz201653g}. But such high temperatures are far beyond the glass transition temperature of the resist and therefore will destroy the patterning mask. Next to thermal annealing also other methods are applied to remove resist residues after the development step of the resist, e.g. low-density inductively coupled Ar plasma \cite{doi:10.1021/nn301093h}, $\text{CO}_2$ cluster cleaning \cite{1.4881635}, or ultraviolet-ozone treatment \cite{1.4754566,1.4868897}. But it still has to be seen if these cleaning methods are able to remove even the last atomic layer of hydrocarbons from the graphene flake. While in most of these publications only rather macroscopic analysis methods such as Raman spectroscopy or even qualitative electrical measurements are used, only the high-resolution X-ray photoelectron spectroscopy in case of the ultraviolet-ozone treatment of reference \cite{1.4868897} may really demonstrate the complete removal of all contaminations. There is, however, also a drawback to this method as long treatments can also create defects in the graphene flake.

\begin{figure}[tb]
\centering
	\includegraphics{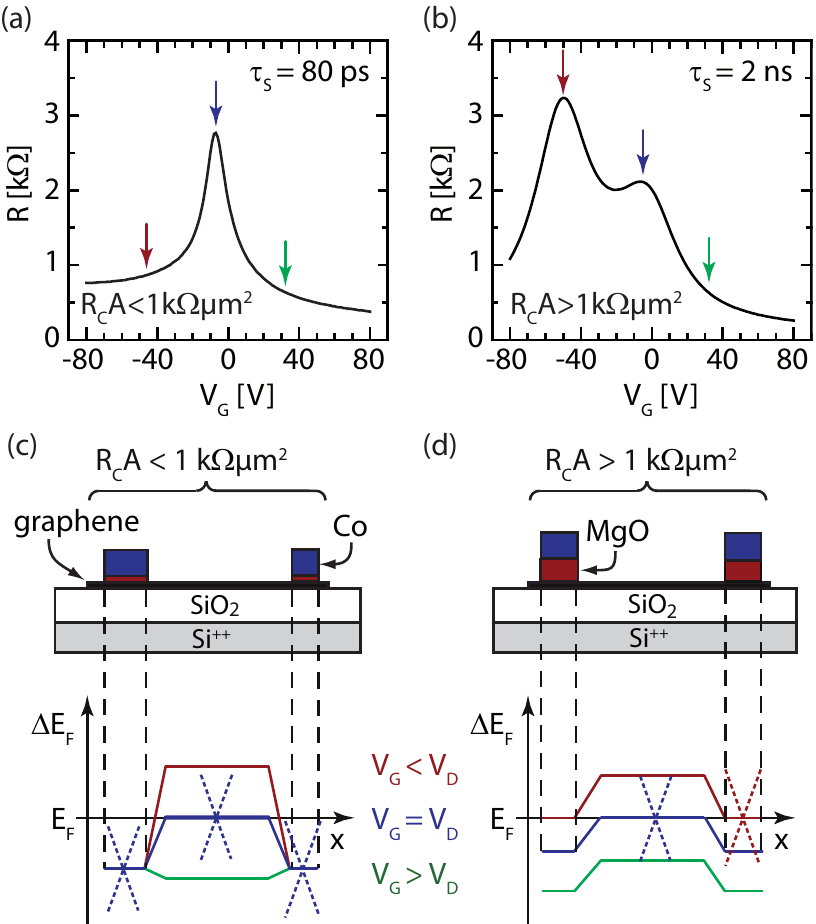}
	\caption{(Color online) (a) and (b) Gate dependent graphene resistance for spin transport devices at room temperature with spin lifetimes of 80~ps and 2~ns (data taken from reference \cite{PhysRevB.90.165403}). Devices with long spin lifetimes exhibit a 2nd charge neutrality point at negative gate voltages (see red arrow in (b)). In addition to a doping of the underlying graphene layer, the interaction of the Co/MgO electrodes with graphene can suppress the electric field effect by the gate voltage $V_{\text{G}}$ which can result in (c) pinning (devices with $R_cA<\unit[1]{k\Omega \mu m^2}$) or (d) no pinning (devices with $R_cA>\unit[1]{k\Omega \mu m^2}$) of the Fermi level in graphene parts underneath the electrodes (corresponding Dirac cones are indicated by dashed lines).}
	\label{figure2}
\end{figure}

The next issue arises because of the chemically inert nature of graphene and its $sp^2$ hybridization, which leads to unfavorable growth conditions of many materials on top of graphene. This can already be seen in the stronger accumulation of the PMMA on the graphene flake compared to the area of SiO$_2$ for dosages lower than $\unit[70]{\mu C / cm^2}$ in figure \ref{figure1}(b) (in this figure the scale is optimized to the height of the underdeveloped PMMA). The reason for this accumulation is the low wettability of graphene \cite{Wang2009} and again it is transmission electron microscopy which reveals the clustering of almost every metal deposited on graphene \cite{doi:10.1021/nl103980h,doi:10.1021/jz201653g}. The interaction between metals and graphene is so weak, that in sub-monolayer growth of metal layers on graphene, the metal atoms solely resides on the hydrocarbon contamination and not on the clean graphene parts \cite{doi:10.1021/nl103980h,doi:10.1021/jz201653g}. For example, the direct growth of MgO on graphene is governed by the Volmer-Weber island growth mechanism \cite{wang:183107} resulting in inhomogeneous oxide layers with presumably pinholes \cite{PhysRevB.90.165403}. This is illustrated in the AFM image of a 3~nm thick MgO layer grown on top of graphene in figure \ref{figure1}(d), which exhibits peak-to-peak values of up to 2~nm. The use of a Ti/TiO$_2$ wetting layer yields more homogeneous MgO layers \cite{wang:183107} but the impact of Ti on the properties of graphene is still an open question \cite{Gong2013a,doi:10.1021/nn502842x}. For Al$_2$O$_3$ the evaporation of Al and its subsequent oxidation to Al$_2$O$_3$ can also yield rough layers with pinholes \cite{Dlubak2012b}. On the other hand, there is a report on homogeneous and pinhole free Al$_2$O$_3$ oxide barriers on graphene by argon sputter deposition \cite{Dlubak2012b}. But the same group also demonstrated by Raman spectroscopy that this technique may also induces defects in graphene \cite{dlubak:092502}. By now, there are only few studies reporting epitaxial-like growth of materials on graphene, e.g. EuO \cite{doi:10.1021/nn303771f} or Ti \cite{doi:10.1021/nn502842x}.

 {The hydrocarbon contaminations from the lithography step and the clustering of many materials on top of graphene are important issues for graphene-based spin transport devices because both can reduce the quality of the insulating oxide barrier between graphene and the ferromagnetic electrodes. But as mentioned above, an insufficient barrier quality can be the bottleneck for the overall spin transport \cite{PhysRevLett.105.167202,PhysRevB.86.235408,PhysRevB.88.161405,PhysRevB.90.165403}. A possible way to circumvent some of these problems was paved by the introduction of mechanical transfer techniques for 2d materials (see e.g. \cite{R.2010,Wang01112013,PRL.99.232104,10.1063.1.4886096,2053-1583-1-1-011002,Neumann_NL2015}).} These transfer techniques provide a controlled deposition of flakes with high spatial precision, which results in two important advantages: (1) The device fabrication becomes now independent of the random position of exfoliated flakes on a substrate and (2) the possibility to deposit different 2d materials on top of each other in order to build so-called van der Waals heterostructures (review about these heterostructures in reference \cite{Geim2013}).

It has been shown that these new transfer techniques can be used to overcome the problem of direct growth of electrode material onto graphene. For this the electrode structure is fabricated by means of e-beam lithography and metallization in a first step. Only afterwards the graphene flake is deposited on top of the prepatterned structure (figure \ref{figure1}(e)) \cite{Droegeler2014}. Therefore, the graphene flake is not exposed to the e-beam lithography step, hence resist residues and e-beam induced defects can be avoided. Furthermore, the growth conditions of the ferromagnetic electrodes and the oxide barrier are far more suitable and can be adapted to an adequate substrate. Especially, the whole expertise already achieved in the fabrication of high quality magnetic tunnel junction devices (e.g. \cite{Parkin2004,Lu2007,Huang2008,Tsunekawa2005,Yang2012,Djayaprawira2005}) may now also be applied to graphene spin transport devices. In this respect, we especially refer to the improvement of magnetic tunnel junctions by annealing which results in crystallization of the interface between ferromagnetic metals and oxide barriers \cite{Lu2007,Yang2012}. The approach to first fabricate spin injection and detection electrodes on a wafer and then to deposit a stack of graphene on hexagonal boron nitride on top of it was e.g. applied in reference \cite{Droegeler2014}, where spin lifetimes of 3.7 ns were measured at room temperature in trilayer graphene, which is the longest room temperature spin lifetime reported so far. {Nevertheless, there might still be some issues with this new fabrication process because the insulating barrier between graphene and ferromagnetic electrode is exposed to air at some point during the process. It is well known that many oxides \cite{Henderson20021} and particularly MgO \cite{Liu1998,Carrasco2010,craft:1507} are hygroscopic. Therefore, it cannot be excluded that a partial hydroxylation of the MgO barrier may still limit the barrier quality in reference \cite{Droegeler2014}.}

As already mentioned, the possibility to deposit and stack various 2d materials on top of each other in order to fabricate van der Waals heterostructures is the second advantage of the transfer techniques. So far, many possible 2d materials beyond graphene have emerged (overview e.g. in \cite{Wang2012c,doi:10.1021/nn400280c,doi:10.1021/cr300263a,Geim2013,doi:10.1021/nn500064s}) and combining these 2d materials with graphene has expanded the field of graphene research dramatically. One of the first applications was the fabrication of stacks consisting of hexagonal boron nitride (hBN) and graphene. Due to the atomically smooth surface and similar lattice constant of hBN to graphene, these devices exhibit significantly improved electrical properties compared to graphene deposited on SiO$_2$ \cite{R.2010,PhysRevLett.113.126801,PhysRevX.4.041019}. The incorporation of such hBN-graphene stacks in graphene spintronic devices led to a significant enhancement of the measured spin diffusion lengths as well as the charge carrier mobility \cite{PhysRevLett.113.086602,Droegeler2014}. Another application of hBN which became feasible with the transfer techniques is its usage as an oxygen-free injection barrier when placed between graphene and the ferromagnetic electrodes \cite{APEX.6.073001,Kamalakar2014,1.4893578}. By transferring exfoliated hBN flakes, the above-mentioned problems in oxide barrier growth may be avoided.

Finally, the transfer techniques offer a new route to functionalize graphene by putting different 2d materials in direct contact to graphene. One important application of such a heterostructure is the enhancement of the spin-orbit coupling (e.g. by the spin-orbit proximity effect of tungsten disulfide on graphene \cite{Avsar2014}) to allow for spin manipulation via the Bychkov-Rashba effect. We emphasize that many effects like the spin-orbit proximity effect greatly rely on contamination-free interfaces. However, depending on the exact kind of transfer technique different amounts of hydrocarbon contamination between the layers of the heterostructures can be observed \cite{Haigh2012,doi:10.1021/nl5006542}.  Interestingly, there seems to be "self-cleansing" effects in some heterostructures (e.g. graphene on hBN, MoS$_2$, and WS$_2$), in which the surface contamination of the 2d materials automatically aggregates into bubbles, leaving behind rather clean interfaces throughout $\mu$m-sized graphene/hBN areas \cite{Haigh2012,doi:10.1021/nl5006542}.

%%%%%%%%%%%%%%
\section{Impact of metals on graphene}
\label{ImpactOnGraphen}

In this section, we discuss the influence of metallic electrodes on both spin and charge properties of the underlying graphene layer. One important tool to investigate the electronic properties is angle-resolved photoemission spectroscopy (ARPES), which can directly probe the band structure of graphene which is in direct contact to the metal (see e.g. references \cite{PhysRevB.82.121101,1367-2630-13-11-113028,Marchenko2012,Varykhalov2012,C2CP42171B} for ARPES measurements of graphene in contact with Au, Ag, Fe, Ni, Co, Al, and Cu). Of special interest in the area of graphene spintronics are the 3d transition metals Fe, Ni, and Co, which are used as ferromagnetic electrodes to generate spin polarized currents. ARPES measurements have revealed that all three metals significantly alter the band structure of graphene: The Dirac point of graphene shifts around 2.8 eV below the Fermi energy and the graphene's $\pi^*$ band hybridizes with 3d bands of Co near the Fermi level \cite{Varykhalov2012,C2CP42171B}.

\begin{figure*}[t]
\centering
	\includegraphics{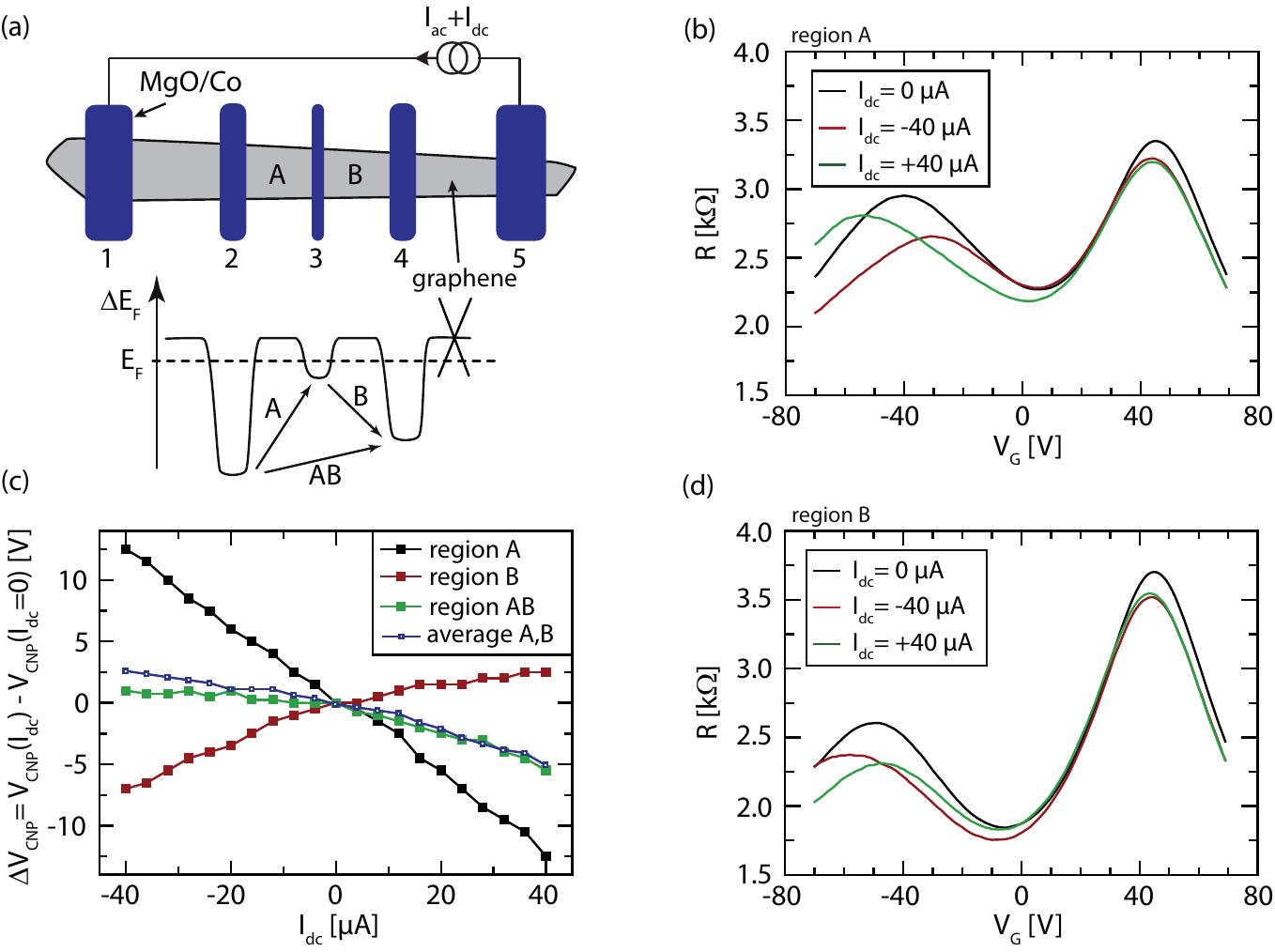}
	\caption{(Color online) (a) Schematic layout of one device with assumed doping profile, which was deduced from the measured contact resistances (see text for more information). (b) and (d) show gate dependent resistance curves of regions A and B as a function of applied dc-current. (c) Gate voltage shift of the contact induced CNP normalized to the gate voltage position without applied dc-current as a function of dc-current.}
	\label{figure3}
\end{figure*}

As a results, as soon as the ferromagnetic electrode gets into direct contact with graphene (which can be the case for pinholes in the separating oxide layer), the spin current is initially, i.e. right after electrical spin injection, no longer carried by pure graphene states near the Dirac point but rather by the hybridized states near the Fermi level. A significant higher spin scattering rate in such states in comparison to the states near the Dirac point of unmodified graphene may explain the overall short spin lifetime measured in devices with ohmic contacts exhibiting small contact resistance area products $R_cA<\unit[1]{k\Omega \mu m^2}$ \cite{PhysRevB.88.161405} (see also figure \ref{figure5}). {Furthermore, the large density of states at the Fermi level due to hybridization results in a pinning of the Fermi level, which is equivalent to a screening of a gate electric field underneath the electrodes \cite{nouchi:253503,H.2008,PhysRevB.79.245430}. Considering both the n-doping and the screening underneath the contacts caused by the Co, the gate voltage dependent doping profiles of figure \ref{figure2}(c) can be deduced for ohmic contacts, which in turn can explain the gate dependent resistance curve of such a device in figure \ref{figure2}(a).}

\begin{figure*}[tb]
\centering
	\includegraphics{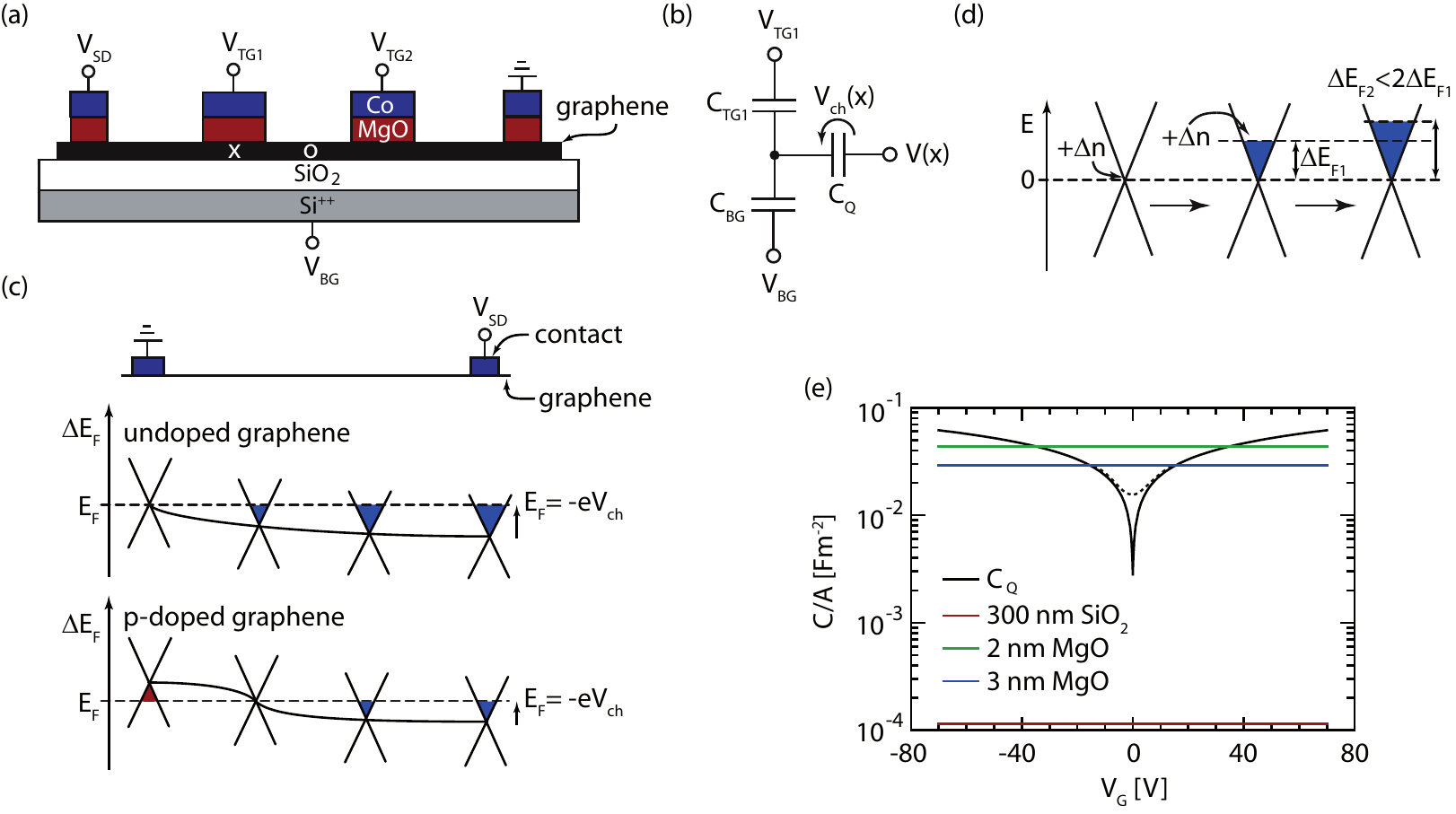}
	\caption{(Color online)(a) Schematic layout of a graphene field effect transistor (see text for the definitions of the applied voltages). In case of the position marked with an X an equivalent circuit is depicted in (b), which can be used to calculate the charge carrier density at this point. (c) Illustration of the Fermi level shift by applying a bias voltage to both undoped and p-doped graphene. (d) Simplified explanation of the quantum capacitance of graphene (see text for more information). (e) Theoretical gate dependence of both the oxide capacitances and the graphene quantum capacitance.}
	\label{figure4}
\end{figure*}

A sufficiently thick oxide barrier without pinholes, which is preferable for spin transport devices, is expected to suppress hybridization between the metal and graphene. Nevertheless, a field effect doping of graphene in such metal/\-dielectric/\-graphene heterostructures is still expected because of different respective work functions \cite{PhysRevLett.101.026803,PhysRevB.79.195425,PhysRevB.87.075414}. The thicker the insulating oxide barrier between graphene and metallic electrode (which results in larger $R_cA$ products) the lower the corresponding doping \cite{PhysRevB.87.075414}. At some point the doping is small enough to be also tuned by the back gate voltage (see corresponding doping profile in figure \ref{figure2}(d)). The different doping of graphene in between and underneath the electrodes can be seen by two distinct charge neutrality points (CNP) in gate dependent resistance measurements (figure \ref{figure2}(b)): the right peak (blue arrow) corresponds to the graphene part between the electrodes (CNP mostly around a gate voltage of $V_\text{G}=\unit[0]{V}$ for untreated devices or at positive gate voltages because of p-doping during oxygen treatments, see reference \cite{PhysRevB.90.165403}) while the left one (red arrow) results from the graphene underneath the electrodes  (CNP at large negative gate voltages because of the n-doping of Co).  {For our Co/\-MgO/\-graphene devices, the contact-induced CNP at negative gate voltage (see red arrow in figure \ref{figure2}(b)) typically appears for $R_cA>\unit[1]{k\Omega \mu m^2}$ \cite{PhysRevB.88.161405,PhysRevB.90.165403}, which is also the value at which the $I$-$V$-curves of the contacts exhibit non-linear behavior \cite{PhysRevB.90.165403}. At this point, we want to emphasize that only a sufficiently thick insulating barrier diminishes the metal-induced doping density of graphene strong enough that the contact-induced 2nd charge neutrality point becomes accessible by large back gate voltages. The required thickness depends on the used contact materials \cite{PhysRevB.87.075414}. Therefore, the value of $R_cA=\unit[1]{k\Omega \mu m^2}$ is most likely only holds for our Co/MgO/graphene devices and may strongly differ for other material combinations.}

In the following, we discuss the origin of the contact-induced second CNP in more detail. Although this review article is about spin transport, we consider this discussion as crucial as it highlights two important aspects which are often neglected: (1) The significant modification of graphene's transport properties right underneath the contacts and (2) the existence of lateral pn-junctions which the spin polarized charge carriers have to overcome.

To confirm that the left CNP in our devices is indeed contact-induced, we now focus on a non-local spin-valve device with Co/MgO electrodes in more detail. This device exhibit a large electrode-to-electrode variation of the respective $R_cA$ values. Figure \ref{figure3}(a) depicts the schematic layout. The measured contact resistance area products of the inner contacts 2 to 4 are: $R_cA_2=\unit[1.2]{k\Omega \mu m^2}$, $R_cA_3=\unit[7.6]{k\Omega \mu m^2}$, and $R_cA_4=\unit[4.8]{k\Omega \mu m^2}$. {According to our previous studies \cite{PhysRevB.88.161405,PhysRevB.90.165403}, we now assume that the interaction between Co and graphene is the dominating source of doping, whereas doping from fabrication-induced contaminations (like resist residues) or from the MgO itself are subordinated. Following the above explanation that thicker barriers with larger $R_cA$ values diminishes Co-induced doping of graphene \cite{PhysRevLett.101.026803,PhysRevB.79.195425,PhysRevB.87.075414}, we can now assign larger doping levels to lower $R_cA$ values.} This yields the potential profile of figure \ref{figure3}(a) for which we assume that the graphene between the contacts is hole doped by an applied gate voltage. The depicted black curve corresponds to the position of graphene's CNP relative to the Fermi level $E_F$ (also see depicted Dirac cone). Gate dependent 4-terminal resistance measurements were performed as a function of dc-current. For this, we apply a total current of $I=I_{\text{ac}}+I_{\text{dc}}=\unit[1]{\mu A}+I_{\text{dc}}$ over the outermost electrodes 1 and 5 and measure the voltage drop between contacts 2 and 3 for region A and contacts 3 and 4 for region B by standard low frequency lock-in technique.

As it can be seen for both regions in figures \ref{figure3}(b) and (d) both the position and magnitude of the left CNP significantly depends on the applied dc-current whereas for the right CNP of the bare graphene there is only a small decrease in resistance for both positive and negative dc current which can be explained by the larger local temperatures from current-induced Joule heating. Figure \ref{figure3}(c) depicts the dependence of the gate voltage position of the contact-induced left CNP as a function of dc-current. We show this data not only for regions A and B but also between contacts 2 and 4, which we call region AB. Interestingly, the gate voltage shift of the CNP has different sign and amplitude in regions A and B. Both, the different sign and amplitude can also be seen in the doping profile of figure \ref{figure3}(a) (indicated by the arrows). {Consistently, the dc-current dependent shift of the contact-induced CNP measured between contacts 2 and 4 (green curve in figure \ref{figure3}(c)) is in good agreement the arithmetic mean of the respective curves of regions A and B (blue curve in figure \ref{figure3}(c)).}

With the help of theoretical considerations about graphene field effect transistors \cite{thiele:094505,thiele:034506,6054021,6018290,PhysRevB.87.125427} we are able to qualitatively understand the gate voltage shift of the contact-induced CNP. For this we assume that high quality tunnel barriers induce a field effect similar to top gates, which changes the Fermi level in graphene underneath the contacts. In fact, a reverse argumentation might be more intuitive (we refer to reference \cite{PhysRevB.87.075414}): Because of the different work functions between graphene and a metal a charge transfer will occur during the alignment of the Fermi levels. The charge transfer yields a carrier doping of graphene, but it also creates a voltage drop and therefore a local electric field over the separating dielectric MgO barrier as the barrier acts as a capacitor with the graphene and the metallic electrodes as its plates. These voltages $V_{\text{TG}i}$ (TG as in top gate) are depicted in the schematic layout of a graphene field effect transistor in figure \ref{figure4}(a). In the following, we assume that these voltages remain unaffected if the two inner contacts in figure \ref{figure4}(a) are used as voltage probes while the outer ones are source and drain contacts (this wiring applies to the measurement of figure \ref{figure3}).

Next, we determine the charge carrier density in the graphene flake. For the graphene underneath an inner electrode (marked by "X" in figure \ref{figure4}(a)) we can apply the equivalent circuit in figure \ref{figure4}(b). The MgO layer, which acts as a barrier for spin injection and spin detection, is approximated to be a top gate with capacitance $C_{\text{TG1}}$, whereas the back gate voltage $V_{\text{BG}}$ is applied over the capacitance $C_{\text{BG}}$, which represents the SiO$_2$ layer. Far more complicated is the incorporation of graphene in this picture. For this we first discuss the Fermi level shift by applying a bias voltage over the graphene flake. In figure \ref{figure4}(c) this effect is illustrated in a very simplified system, in which e.g. the doping effects of the contacts are neglected. We first consider ideal, undoped graphene, in which case the Fermi level of the graphene underneath the grounded contact (left contact in figure \ref{figure4}(c)) is at the Dirac point. A bias voltage applied to the right contact results in a lateral voltage gradient along the graphene channel with corresponding shifts of the Fermi level \cite{thiele:094505,6054021}. Along the whole graphene flake a gradual nn'-junction is generated, which can be verified by photocurrent measurements \cite{Freitag2013}.

The Fermi level shift underneath the contacts with the applied source-drain-voltage $V_{\text{SD}}$ can be written as $E_{\text{F}}\equiv -eV_{\text{ch}}$ where the energy scale is set to zero at the CNP of graphene. In this simplified system the voltage drop $V_{\text{ch}}$ along the graphene channel only depends on the electrostatic potential by the applied bias. But in general, the Fermi level also depends on the electrochemical potential. Accordingly, the position of the Fermi level $E_{\text{F}}= -eV_{\text{ch}}$ also depends on other factors such as the applied gate voltage or changes in the electrochemical potential by adsorbates. This is also illustrated in figure \ref{figure4}(c) in the case that graphene is p-doped by an applied gate voltage or adsorbates. Apparently, the local area with vanishing charge carrier density, i.e. where the Fermi level is at the Dirac point, moves along the graphene channel (from left to right in the lower panel of figure \ref{figure4}(c)) with increasing p-doping concentration. Such a shift of the CNP by an applied gate voltage is e.g. observed in photocurrent and thermal infrared microscopy \cite{Freitag2013,Loomis2012,Freitag2010,Bae2010}.

The capacitance of the graphene part in the equivalent circuit over which the voltage $V_{\text{ch}}$ drops is the quantum capacitance $C_{\text{Q}}$ of graphene \cite{luryi:501,fang:092109}. The quantum capacitance directly results from the Pauli principle and becomes relevant in materials with a small density of states (DOS) near the Fermi level. To explain the quantum capacitance, we first consider ideal graphene with the Fermi level at the Dirac point, where the DOS vanishes (figure \ref{figure4}(d), left band structure).  If now the charge carrier density in graphene is changed by $\Delta n$ the Fermi level shifts by $\Delta E_{\text{F1}}$ as the electrons have to occupy higher energy states. For single layer graphene, the Fermi level $E_\text{F}$ and the DOS at the Fermi level $D(E_{\text{F}})$ are given by \cite{RevModPhys.83.407}:
\begin{equation}
E_{\text{F}} = \hbar v_{\text{F}} \sqrt{\pi n} \quad,\quad D(E_{\text{F}}) = \frac{2 E_{\text{F}}}{\pi \left(  \hbar v_{\text{F}}  \right)^2} \;,
\label{eq:energy}
\end{equation}
where $v_{\text{F}}$ is the Fermi velocity. If now the charge carrier density is doubled by adding the same amount of charges $\Delta n$ a second time, the increase of the Fermi level becomes less strong because of the previous increase in DOS ($\Delta E_{\text{F2}}<2\Delta E_{\text{F1}}$) (see figure \ref{figure4}(d)). As mentioned above, the electrons at the Fermi level have the energy $E_{\text{F}} = -eV_{\text{ch}}$. Combining this expression with equation \eqref{eq:energy} and the general definition of a capacitance $C_\text{Q}=\partial Q / \partial V_{\text{ch}}$ (with the charge $Q=Ne=Ane$ and area $A$) results in the quantum capacitance of graphene:
\begin{equation}
\frac{C_{\text{Q}}}{A}=\frac{2}{\pi}\frac{e^3 V_{\text{ch}}}{ \left( \hbar v_{\text{F}} \right)^2} =\frac{2}{\pi} \left( \frac{e}{ \hbar v_{\text{F}}} \right)^2 E_{\text{F}} = \frac{2}{\sqrt{\pi}} \frac{e^2}{\hbar v_{\text{F}}} \sqrt{n}\;.
\end{equation}

In figure \ref{figure4}(e) we plot this quantum capacitance of graphene as a function of the applied gate voltage (black solid line). Due to broadening of the DOS by both thermal excitation at room temperature and by structural inhomogeneities the experimentally determined graphene quantum capacitance is also broadened \cite{Xu2011,Xia2009a} which is depicted as a black dashed line. Furthermore, we include the capacitance of both oxide layers (the MgO of the top electrode with thicknesses of 2 and 3~nm and the SiO$_2$ of the back gate).

Now we go back to the equivalent circuit in figure \ref{figure4}(b). First, we start with the graphene part between the contacts (marked by a circle in figure \ref{figure4}(a)). As there is no top gate capacitance in this region, back gate and quantum capacitances are put in series and give a total capacitance $C_\text{total}=C_\text{BG} C_\text{Q}/(C_\text{BG}+C_\text{Q})$. As the back gate capacitance is much smaller than the quantum capacitance (compare to figure \ref{figure4}(e)), it dominates the total capacitance: $C_\text{total}\approx C_\text{BG}$. Accordingly, the quantum capacitance does not play any significant role in the back gate induced field effect (see also \cite{luryi:501,fang:092109}). This is an important conclusion as the bias voltage can only change the charge carrier density at the node of the equivalent circuit in figure \ref{figure4}(b) with help of the quantum capacitance. The fact that the quantum capacitance can be neglected now explains why the CNP of the bare graphene part between the contacts does not shift with dc currents (figures \ref{figure3}(b) and (d)).
In contrast, the quantum capacitance plays a significant role for graphene parts underneath the electrodes as it is in the same order of magnitude as the electrode capacitance (figure \ref{figure4}(e)). If we assume a fixed charge carrier density at the node of the equivalent circuit of figure \ref{figure4}(b) for the contact region, now the shift of the contact-induced CNP in figures \ref{figure3}(b) and (d) becomes clear: By changing the dc current (or accordingly the bias voltage) the voltage $V_{\text{ch}}$ will change according to the discussion in figure \ref{figure4}(c). If we want to keep the charge carrier density at the node of the equivalent circuit at $n=0$ (i.e. to the contact-induced CNP) the voltages $V_{\text{TG1}}$ (i.e. the actual doping of the graphene by the contacts) and $V_{\text{BG}}$ also have to change. To simulate this, all changes of $V_{\text{TG1}}$, $V_{\text{BG}}$, and $V_{\text{ch}}$ have to be calculated self-consistently (e.g. a change in the back gate voltage will shift the Fermi level and therefore the value of the graphene quantum capacitance), which goes beyond the scope of this paper. For a far more detailed description we refer to the references \cite{thiele:094505,thiele:034506,6054021,6018290,PhysRevB.87.125427}.

Finally, we again point to the fact that the different charge transport properties of graphene underneath the contacts and graphene in between the contacts are so far widely neglected. It has to be seen how the change in the doping underneath the contacts with different dc-biases or the pn-junctions along the graphene channel have an impact on the performance of graphene-based spintronic devices \cite{PhysRevB.79.081402,PhysRevLett.102.137205}.

%%%%%%%%%%%%%%
\section{Determination of spin lifetimes}
\label{lifetime}

\begin{figure}[tb]
\centering
	\includegraphics{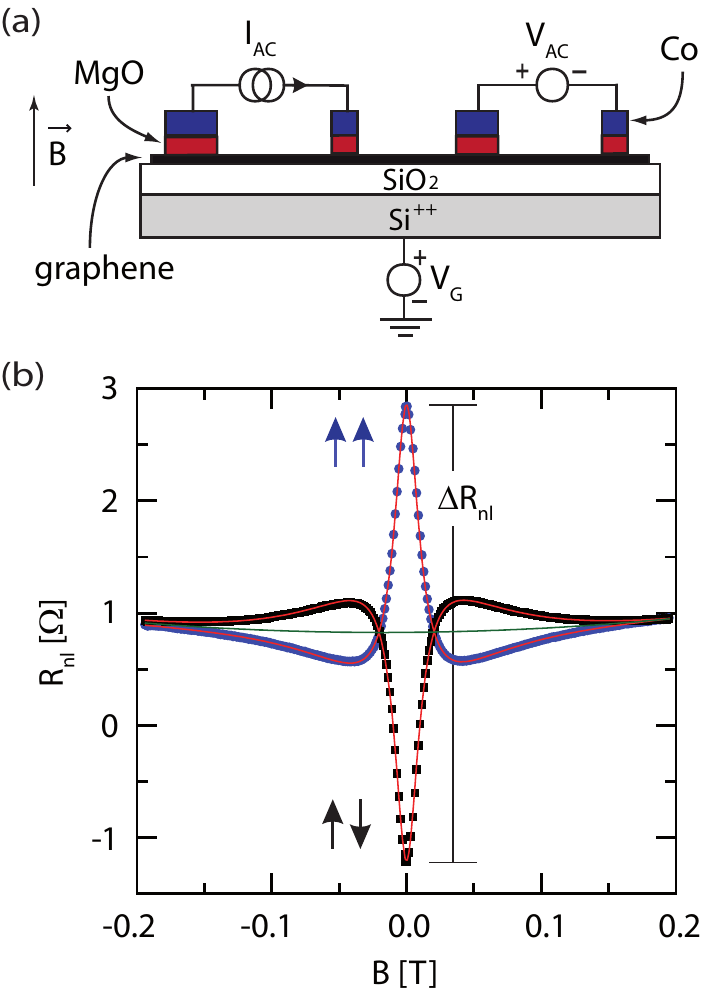}
	\caption{(Color online) (a) Schematic cross section with wiring configuration for non-local spin-transport measurements. (b) Hanle spin precession measurement of a bilayer graphene device for a perpendicular magnetic field sweep with parallel and antiparallel alignments of the respective spin injection and detection electrodes}
	\label{figure8}
\end{figure}

In this section we discuss on how to extract the spin lifetime out of spin precession measurements and address how several limitations to this approach have recently been recognized. We restrict ourselves to non-local spin transport measurements, so we can neglect the drift term in the Bloch-Torrey equation and only consider changes in the net spin vector $\vec{s}$ by spin precession about a perpendicular magnetic field $B$ (see measurement configuration in figure \ref{figure8}(a)), spin diffusion, and spin dephasing and relaxation \cite{RevModPhys.76.323}:
\begin{equation} \label{eq:Bloch}
	\frac{\partial \vec{s}}{\partial t}\;=\;\vec{s}\times \vec{\omega}_{\,0} +D_{\text{s}}\nabla^2\vec{s}-\frac{\vec{s}}{\tau_{\text{s}}}\;.
\end{equation}
Here $\vec{\omega_0}=g\mu_{B} \vec{B}/\hbar$ is the Larmor frequency, $g$ is the gyromagnetic ratio, $D_{\text{s}}$ the spin diffusion constant, and $\tau_{\text{s}}$ the spin lifetime.

The advantage of non-local measurements is that the spin signal can be significantly decoupled from spurious charge signals \cite{PhysRevB.37.5312,RevModPhys.76.323,ISI000249789600001}. To achieve such a non-local configuration, dc measurements or lock-in techniques with low frequencies are needed, because electric pulses in case of RF measurements propagate through the whole device, which means that no non-local part can exist. The dc or low frequency measurements lead to a stationary or quasi-stationary condition and therefore the time derivative of equation \eqref{eq:Bloch} is set to zero. Hence a solution to
\begin{equation} \label{eq:BlochStationary}
	\vec{s}\times \vec{\omega}_{\,0} +D_{\text{s}}\nabla^2\vec{s}-\frac{\vec{s}}{\tau_{\text{s}}}\;=0.
\end{equation}
has to be found to extract the spin lifetime by means of a fit to the dc spin precession curve (so-called Hanle curve). Typical Hanle curves for a spin transport device are shown in figure \ref{figure8}(b) with both parallel and antiparallel alignments of the inner Co electrodes in figure \ref{figure8}(a). The non-local spin resistance $\Delta R_{\text{nl}}$ can be determined at $B=0$. As expected, both Hanle curves merge at larger magnetic fields. However, they do not become constant but rather increase above $|B|>0.2$~T. We note that this magnetic field dependent background signal is typical for most studies and it can even be much more pronounced. {One explanation for this background is that the magnetization of the ferromagnetic electrodes can rotate out-of-plane with increasing perpendicular magnetic field \cite{Nature.416.713-716,PhysRevLett.101.046601,Idzuchi2012,PhysRevLett.113.086602}. In principle, this rotation of the magnetization can account for a background signal which is symmetric in magnetic field. But we often observe an antisymmetric background signal with a linear term in $B$, which also dependents on the charge carrier density and the wiring of the device \cite{2053-1583-2-2-024001}. These findings cannot be explained by the rotation of the electrode's magnetization alone. We have recently shown that such a background of second polynomial order can be caused by an inhomogeneous current flow through the oxide barriers. As a result there is a charge accumulation signal next to the actual spin accumulation signal in the non-local voltage which can be explained by a redistribution of charge carriers by a perpendicular magnetic field similar to the classical Hall effect \cite{2053-1583-2-2-024001}.}

One fundamental drawback of every solution to equation \eqref{eq:BlochStationary} is that the three parameters $g$, $D_{\text{s}}$, and $\tau_{\text{s}}$ cannot be determined independently. This is because of the steady state condition, which allows the multiplication of equation \eqref{eq:BlochStationary} by any factor $\alpha$ without changing the overall result. Therefore a fit using a solution to equation \eqref{eq:BlochStationary} is invariant with respect to the transition ($g, D_{\text{s}}, \tau_{\text{s}})\rightarrow(\alpha g, \alpha D_{\text{s}},\tau_{\text{s}}/\alpha )$. Hence one of the three parameters has to be assumed or determined by other measurements.

Often $g=2$ is assumed for the analysis of spin precession measurements. For pristine or moderately modified graphene flakes this assumption is confirmed by electron spin resonance (ESR) measurements \cite{doi:10.1021/nn302745x,AugustyniakJablokow2013118,Singamaneni2014}. It remains to be seen if this value also holds for functionalized graphene. By now, spin precession measurements of hydrogenated graphene suggest larger $g$-factors \cite{PhysRevLett.109.186604,1.4803843,PhysRevB.87.081402}. Of course, the most elegant way to circumvent this uncertainty is to give up on the non-local, steady state condition and apply time-resolved RF measurements which offer the possibility to directly determine the $g$-factor from time dependent spin precession.

\begin{figure}[tb]
\centering
	\includegraphics{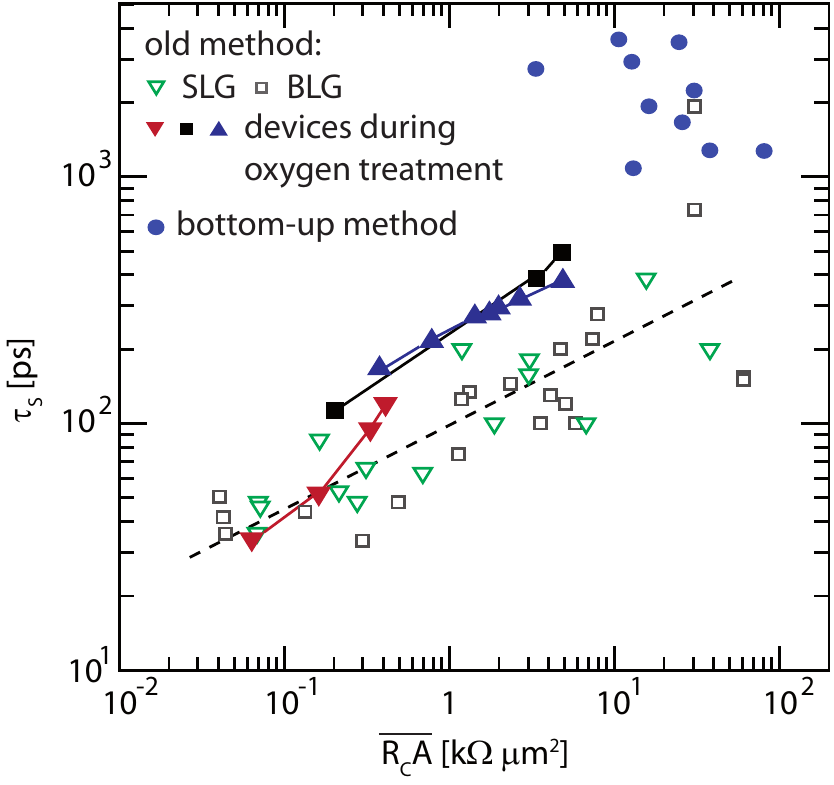}
	\caption{(Color online) Spin lifetime vs averaged contact resistance area product $\overline{R_cA}$ of respective injection and detection electrode at an electron density of $n=\unit[1.5 \times
10^{12}]{1/cm^2}$ at room temperature. {Shown are data from our old fabrication method (as-fabricated single and bilayer graphene devices from reference \cite{PhysRevB.88.161405}) and three devices where the contact resistances were successive increased by oxygen treatments from reference \cite{PhysRevB.90.165403}) as well as data from our new bottom-up fabrication method from reference \cite{Droegeler2014}.}}
	\label{figure5}
\end{figure}

In the following, we discuss some assumptions and simplifications, which have to be made in order to find useable solutions to equation \eqref{eq:BlochStationary}. One of the first and still widely used analytical solution was developed by Johnson and Silsbee \cite{PhysRevB.37.5312}. But in recent years it was found that the measured spin lifetime of a graphene device scales with the $R_cA$ product at least up to values of several tens of k$\Omega\mu$m$^2$ (see figure \ref{figure5}) \cite{PhysRevLett.105.167202,PhysRevB.88.161405,PhysRevB.90.165403,Kamalakar2014}) suggesting that the measured spin lifetime is not the intrinsic spin lifetime of graphene but rather has extrinsic origin. One route to explain this dependence is that the measured spin lifetime can be significantly underestimated if spin relaxation by the contacts is not included in the fit model and hence some work was done in the direction of more elaborated models \cite{PhysRevB.86.235408,PhysRevB.89.245436, 2014arXiv1411.2949I}. But even these models make two crucial assumptions: (1) The injection barriers are homogeneous and can be characterized by the $R_cA$ value only and (2) the spin lifetime in graphene underneath the contacts is the same as the spin lifetime in the bare graphene part between the contacts.

As it was already discussed in section \ref{ImpactOnGraphen} the growth of homogeneous oxide barriers on graphene is quite challenging. But as long as the contact is not spatially homogenous, the measured contact resistance can only be an averaged value. Especially if there are pinholes within the barrier the contact resistance area product cannot precisely be determined, because the exact number and sizes of the pinholes cannot be deduced from electrical measurements. Therefore, the uncertainty in the contact resistance will also lead to an uncertainty in the extracted spin lifetimes by the aforementioned models.

And also the second assumption of a single spin lifetime seems to be oversimplified considering section \ref{ImpactOnGraphen}. As it was demonstrated by ARPES measurements \cite{Varykhalov2012,C2CP42171B} the interaction between graphene and ferromagnetic metals can significantly change the band structure of graphene due to hybridization. The hybridized states are directly at the Fermi level and therefore both charge and spin transport will occur through these states, which may have a strong impact on the spin lifetime. But even if we assume a sufficiently thick oxide barrier, which inhibits this hybridization, the assumption of a single spin lifetime may still be too simple if the spin lifetime depends on the Fermi level or, correspondingly, the charge carrier density. Because the field effect doping of the contacts \cite{PhysRevB.87.075414} can e.g. lead to the case at which the graphene underneath the contacts is highly n-doped whereas the graphene between the contacts is at its CNP. By now, only some research was done on the topic how two spatially different graphene parts with different spin relaxation times influence the overall measured spin lifetime (e.g. in reference \cite{doi:10.1021/nl301050a}).

Finally, we note that a rough MgO layer like the one in figure \ref{figure1}(d) not only can cause pinholes but also leads to an equally rough surface of the Co electrode which is deposited on top. But such a rough ferromagnetic interface can yield stray fields, which may result in a broadening of the spin precession curve and thereby an apparent reduction of the extracted spin lifetime \cite{PhysRevB.84.054410}.

%%%%%%%%%%%%%%
\section{Determination of charge carrier mobilities}
\label{Mobilities}

\begin{figure*}[tb]
\centering
	\includegraphics{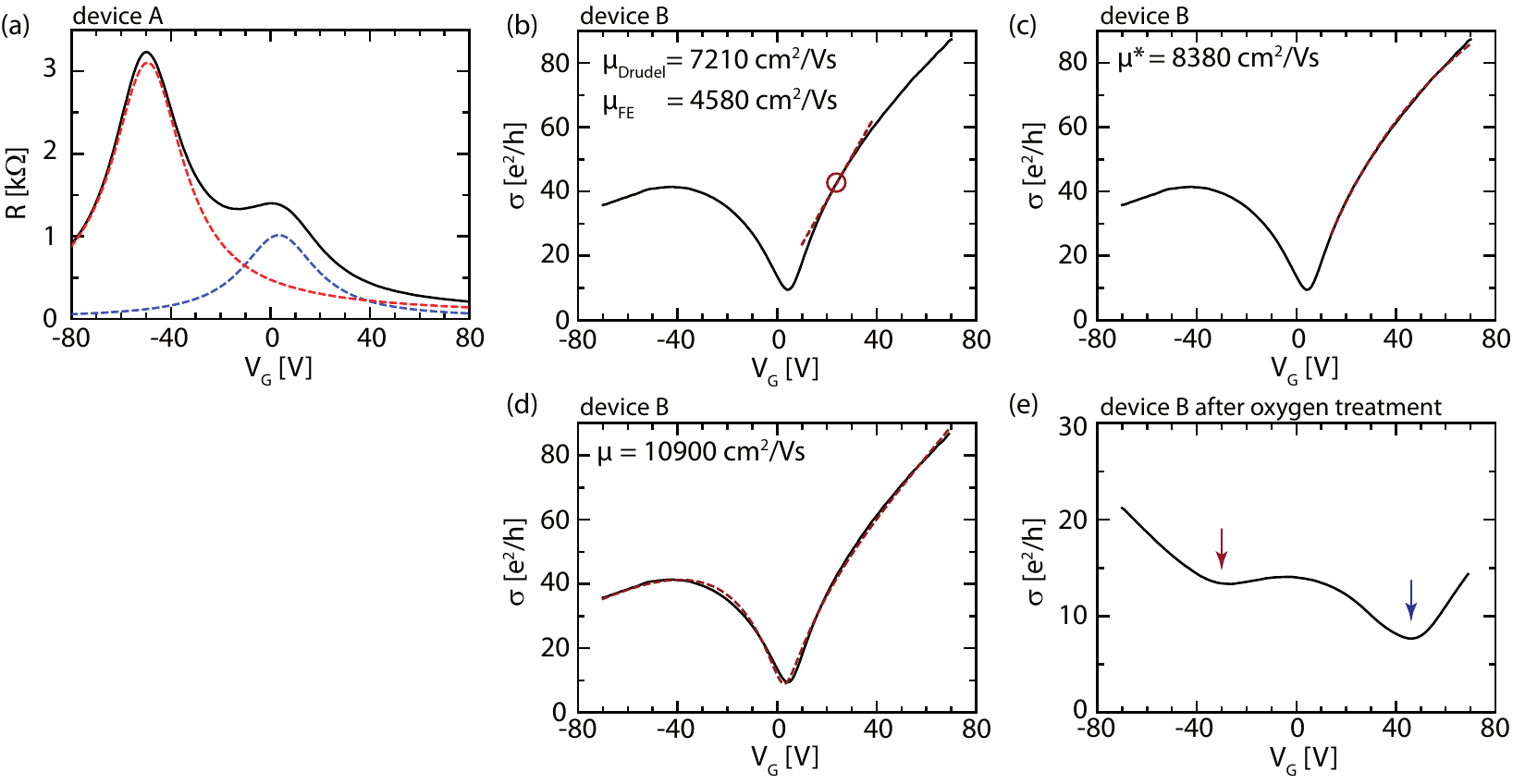}
	\caption{(Color online) {(a) Decomposition of a gate dependent resistance measurement into two Dirac curves for device A corresponding to the n-doped graphene underneath the contacts and the undoped graphene between the contacts. Figures (b)-(d) show the conductivity curve of device B, which is fitted with different models for the determination of the charge carrier mobility (dashed lines). See text for more information. (e) After the contacts of the same device B were oxidized, the contact induced CNP shifted into the measurable gate voltage range.}}
	\label{figure6}
\end{figure*}

In this section we discuss the determination of the charge carrier mobility $\mu$ from gate dependent resistance measurements. There are two reasons making this an important topic for graphene-based spin transport studies. Firstly, the dependence of the spin lifetime and the charge carrier mobility is often used to identify the dominating spin relaxation mechanism \cite{RevModPhys.76.323,ISI000249789600001}. Secondly, the use of oxide barriers for spin injection and detection can lead to the appearance of the contact-induced CNP as it has been discussed in section \ref{ImpactOnGraphen}. We will demonstrate that this 2nd CNP has significant impact on the determination of the charge carrier mobility.

Firstly, we analyze the electron mobility in one of our devices by different methods which are currently used in literature showing that the determination of the mobility is not unambiguous and yield values which vary by more than a factor of 2. The simplest model considers a graphene field effect transistor, for which a linear dependence between the charge carrier density $n$ (or accordingly back gate voltage) and the conductivity $\sigma$ is often observed (see e.g. \cite{Novoselov2004}). For this case several groups (e.g. \cite{PhysRevLett.99.246803,Bolotin2008351,PhysRevLett.102.236805}) assume a simple Drude model and define the Drude mobility as:

\begin{equation}
\mu_{\text{Drude}} = \sigma/(n e).
\label{eq:local}
\end{equation}

We emphasize that this Drude mobility is extracted at a single point of the conductivity curve. To account for the residual conductivity $\sigma_{\text{0}}$ even in the case of $n\rightarrow 0$ the conductivity can be written as $\sigma = \mu e n + \sigma_{\text{0}}$. Starting from this equation the extracted field effect mobility $\mu_{\text{FE}}$ from fitting the slope of the conductivity curves is given by (e.g. in references
\cite{doi:10.1021/nl301567n,PhysRevLett.107.047206}):

\begin{equation}
\mu_{\text{FE}}=(1/e)(\Delta \sigma/\Delta n).
\label{eq:slope}
\end{equation}

We applied both methods to a conductivity curve from one of our actual spin valve devices in figure \ref{figure6}(b). The red circle marks the point at which $\mu_{\text{Drude}}$ and $\mu_{\text{FE}}$ have been determined. Remarkably, $\mu_{\text{Drude}}$ exceeds $\mu_{\text{FE}}$ by almost 60$\%$.
In the as-fabricated state of this device the conductivity drops only slightly at large negative gate voltages, which indicates that the contact induced CNP is close but nevertheless out of range of the applicable gate voltage. After an oxygen treatment which primarily reduces the doping effect of the electrodes \cite{PhysRevB.90.165403} the contact-induced CNP shifts into the accessible gate voltage range (figure \ref{figure6}(e)) and furthermore complicates the analysis of the carrier mobilities.

This can be seen in the resistance curve of another spin valve device in figure \ref{figure6}(a) that also shows two distinct CNPs. We can decompose resistance contributions of both graphene regions (underneath and in between the contacts; model for fitting is discussed further below), which is shown by red and blue dashed lines, respectively. The overlapping of both curves clearly demonstrates one major issue of the two mobility fits discussed so far: As soon as the doping underneath the contact approaches the one of the bare graphene part, both the absolute value and the slope of the measured gate dependent resistance curve significantly changes in the overlapping part. And so does the extracted mobility. The reason for this is quite clear: Both equation \eqref{eq:local} and \eqref{eq:slope} are based on one well-defined transport regime, whereas the contact-induced CNP points to the fact that the charge transport in graphene underneath and in between the contacts can differ significantly.

But before we discuss more detailed fitting models, which include different transport regimes, we first want to discuss another issue when using equation \eqref{eq:slope} which arises as soon as the conductivity curves are not linear as the ones in figure \ref{figure6}. In this case, the extracted mobility $\mu$, which is determined by the slope, depend on the charge carrier density $n$. But this is a mathematical contradiction which is often neglected. The derivative of $\sigma$ with respect to $n$ is now given by:

\begin{align}
	\frac{\text{d}\sigma}{\text{d}n} &=\frac{\text{d}}{\text{d}n} \left( ne\mu (n) + \sigma_{\text{0}} \right) = e \left( \mu (n) + n \frac{\text{d} \mu (n)}{\text{d} n} \right) \\
	\Rightarrow \mu (n) &=\frac{1}{e} \left( \frac{\text{d}\sigma}{\text{d}n}-n\frac{\text{d} \mu (n)}{\text{d} n} \right).
\end{align}

The term $\text{d} \mu (n)$/$\text{d} n$ is neglected by the approach of equation \eqref{eq:slope} and only vanishes for strictly linear conductivity curves.

But the question arises if the mobility indeed depends on the charge carrier density in case of a non-linear conductivity curve as the non-linearity can also be explained by a constant mobility ($\text{d} \mu$/$\text{d} n=0$) if there is a charge carrier independent contribution $\rho_0$ to the overall resistivity. Such a contribution can result from both short- and correlated long-range disorder \cite{PhysRevLett.107.156601}. Hence, the corresponding conductivity is given by (see e.g. references \cite{PRL.99.232104,PhysRevLett.107.206601}):

	\begin{equation}
	\frac{1}{\sigma}=\frac{1}{ne\mu^*+\sigma_0}+\rho_0.
\label{eq:partial}
\end{equation}

Figure \ref{figure6}(c) demonstrates that a significant part of the electron branch for $V_{\text{G}}>$~0 of the non-linear conductivity curve can be fitted under the assumption of the constant mobility $\mu^*=8380$cm$^2$/Vs. However, the fitting fails completely in the hole branch (not shown), because here the impact of the contact-induced CNP has to be considered. Furthermore, all three approaches that are discussed so far break down near the CNP. But there are also models trying to fit the whole conductivity curve including the two CNPs (e.g. in reference \cite{Vera-Marun2012}). We also use the following model which assumes two differently doped graphene parts with different mobilities $\mu_{\text{i}}$:
\begin{equation}
	\frac{1}{\sigma}=\sum_{i=1,2} \frac{1}{\sqrt{(n_i^2+n_{i0}^2)} \; e\mu_{\text{i}}}+\rho_0,
	\label{eq:global}
\end{equation}
where $n_i= \alpha (V_{\text{G}}-V_{\text{CNP,}i})$ is the gate voltage $V_{\text{G}}$ dependent charge carrier density, $\alpha=\unit[7.18 \times
10^{10}]{1/(cm^2V)}$ the capacitive coupling for a $\unit[300]{nm}$ thick SiO$_2$ layer, $n_{i0}$ the residual charge carrier densities due to the presence of electron-hole puddles \cite{Martin2008} and thermally excited carriers, which prevent a divergence at the CNPs \cite{Vera-Marun2012}, and a gate independent resistivity $\rho_0$. This model can be fitted quite well to the conductivity curve of figure \ref{figure6}(d) (the given mobility $\mu=10900$~cm$^2$/Vs corresponds to the Dirac curve with its CNP near zero gate voltage).

But now there are two issues: At first, the values of the charge carrier mobilities as extracted from the four models spread over a broad range between $\unit[4580]{cm^2/(Vs)}$ and $\unit[10900]{cm^2/(Vs)}$. Secondly, there is a serious issue when extracting the mobility from equation \eqref{eq:global}, which already contains seven parameters, as it cannot be used to adequately fit the conductivity curve of the same device after oxygen treatment (figure \ref{figure6}(e)). We tried several other models to fit the curve in figure \ref{figure6}(e), but for a reasonable fit we need at least eight free parameters. But as we already briefly discussed in one of our previous publications \cite{PhysRevB.88.161405} and will discuss in this section in more detail, such a large number of free parameters make such a model useless. Because as long as the Dirac curves have symmetric shapes, which they have in figure \ref{figure6}(a) and in our previous publications \cite{PhysRevB.88.161405,PhysRevB.90.165403}, only seven parameters can reliably be extracted out of a gate dependent resistivity curve with two visible CNPs: The position, magnitude and width for each of the two peaks and the background. Therefore, a model of eight parameters is overdetermined and accordingly we are able to change the mobility values in such a model over an unreasonable wide range without seriously worsening the fit results.

To approach reliable mobility fits, we now deduce several parameters which have to be included into a conductivity model to adequately fit a gate dependent resistivity curve with two CNPs. First, there is the carrier mobility as the actual quantity of interest. As already discussed in section \ref{ImpactOnGraphen}, the interaction with the electrodes can significantly change the electronic properties of graphene. We therefore have to consider two mobilities for each device: one for the graphene underneath the contacts and one for the graphene area in between the contacts. The same argument holds for the electrochemical doping of the respective graphene parts, which is obvious as we observe two CNPs. Two additional parameters are needed to include spatial variations of the electrochemical potentials within each graphene region. These two parameters can also account for the minimum conductivities as the variations in the potential lead to the electron-hole puddles at the charge neutrality point \cite{Vera-Marun2012}. So far this yields six parameters.

\begin{figure*}[!tb]
\centering
	\includegraphics{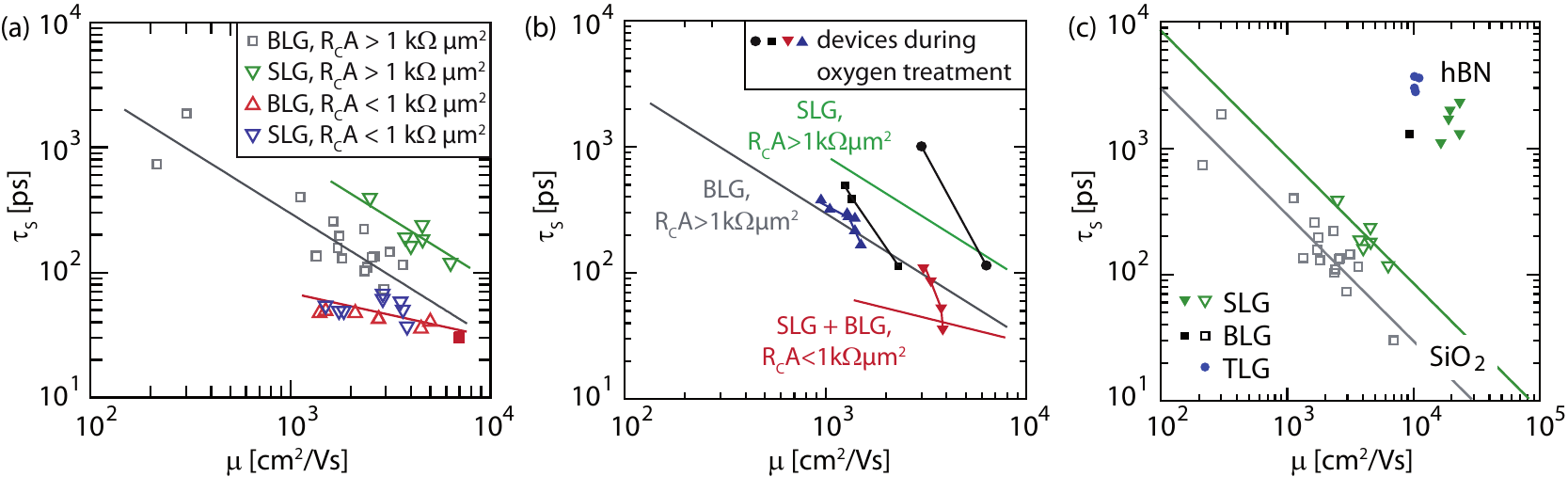}
	\caption{(Color online) (a) Spin lifetime vs electron mobility at $n=\unit[1.5 \times 10^{12}]{1/cm^2}$ at room temperature for as-fabricated single and bilayer graphene devices. Only for devices with $R_cA>\unit[1]{k\Omega \mu m^2}$ a D'yakonov-Perel' like $1/\mu$-dependence can be fitted to the data. The line for the devices with $R_cA<\unit[1]{k\Omega \mu m^2}$ is just a guide to the eye and illustrates the deviation from a $1/\mu$-dependence. (b) Same trends as in figure (a) combined with the development of the spin lifetime and mobility of four devices during oxygen treatment. (c) Devices with $R_cA>\unit[1]{k\Omega \mu m^2}$ both fabricated with our old process (exfoliated graphene on SiO$_2$ with subsequent deposition of the electrodes) and the new one (prepatterned electrodes and transfer of a hBN/graphene-stack like depicted in figure \ref{figure1}(e)). Data taken from references \cite{PhysRevB.88.161405,PhysRevB.90.165403,Droegeler2014}.}
	\label{figure7}
\end{figure*}

In the next step, we have to critically review the gate voltage as the tuning parameter for the resistivity measurement. As discussed e.g. in \cite{nouchi:253503,H.2008,PhysRevB.79.245430} the interaction between contact and graphene may lead to a screening of the gate electric field underneath the electrodes. We assume that there is a gradual transition from pinning to depinning of the Fermi level underneath the contacts with increasing oxide barrier thickness and quality. Therefore, we have to consider a screening factor as a seventh parameter which accounts for the effective Fermi level shift in the graphene underneath the contacts as a function of applied gate voltage.

Furthermore, there is the unknown transition of the electrochemical potentials between contact covered and bare graphene part, i.e. the exact shape and lateral extension of the pn-junctions, which are known to exist near the edges of the contacts \cite{H.2008,PhysRevB.79.245430}. In the most simple approximation, at least one more parameter has to be considered which describes the decaying length of the pn-junction (see e.g. references \cite{JJAP.50.070109,PhysRevB.82.115437}). And finally, there might be a gate voltage independent contribution $\rho_0$ to the overall resistance as already discussed in equation \eqref{eq:partial}. Hence, at least nine parameters are necessary to simulate the gate dependent resistivity, whereas only seven independent parameters may be experimentally determined by a gate dependent resistance measurement.

The whole problem with the analysis of the carrier mobility from gate dependent resistance measurements is that both the contact-covered and bare graphene parts are measured simultaneously in series. The lack of independently probing both regions is bound to the very nature of a pure electrical transport measurement. Therefore, independent measurements have to be conducted to determine some of the aforementioned parameters separately. Especially, information about the spatial change in the electrochemical potential along the contact induced pn-junctions may be helpful. These can be obtained from scanning photocurrent microscopy \cite{H.2008,PhysRevB.79.245430,Park2009}.

Finally, we note that most other graphene-based nanoelectronic devices are fabricated with low-ohmic, metallic contacts. This might be the reason why the aforementioned issues concerning the determination of the charge carrier mobility is not discussed in recent publications in more detail. For metallic contacts, a complete pinning of the Fermi level underneath the contacts can be assumed (see section \ref{ImpactOnGraphen}). Therefore, the complete contact area of the device only contributes as a gate independent constant resistance to the overall gate dependent resistance (e.g. $\rho_0$ in equation \eqref{eq:partial}). Additionally, the impact of both the contact area and the pn-junctions on the transport measurement decreases with increasing separation of the electrodes. We note that in many charge-based electrical transport studies the contact separation is much larger than in graphene-based spin transport devices.

%%%%%%%%%%%%%%
\section{{Critical review and perspectives}}
\label{SpinRelaxationMechanism}

{In this final section, we critically review some experimental studies which were used to investigate the dominant spin relaxation mechanism in graphene.} Maybe the most direct way to identify the dominating spin relaxation mechanism is to evaluate the dependence of the spin lifetime on the charge carrier mobility \cite{RevModPhys.76.323,ISI000249789600001}. A linear dependence of $\tau_s$ on $\mu$ or $\tau_p$, which is the momentum scattering time, is a priori suggestive of an Elliott-Yafet (EY) spin scattering mechanism, while the inverse relation $\tau_s \propto 1/\mu \propto 1/\tau_p$ indicate the dominance of a D'yakonov-Perel'-like (DP) spin scattering mechanism. In our previous studies we found an inverse dependence of spin lifetime on the mobility for both single-layer \cite{PhysRevB.88.161405} and bilayer \cite{PhysRevLett.107.047206} graphene which is thus indicative for DP-like spin scattering. This dependence was only found in devices with contact resistance area products larger than $R_cA>\unit[1]{k\Omega \mu m^2}$ (figure \ref{figure7}(a)). In contrast, all devices with $R_cA<\unit[1]{k\Omega \mu m^2}$ (figure \ref{figure7}(a)) and devices where the contact resistances were enhanced by subsequent oxygen treatments (figure \ref{figure7}(b)) do not show this 1/$\mu$-dependence.
All these devices have been prepared by the conventional top-down method for which the MgO injection and detection barrier was directly deposited onto graphene.

Most strikingly, the new generation of devices where we apply the bottom-up approach by transferring a hBN-graphene-stack on prepatterned electrodes (see figure \ref{figure1}(e) and section \ref{NewMethods}) exhibit significantly enhanced charge and spin transport properties (see full symbols in figure \ref{figure7}(c)). We attribute the increase in mobility to the hBN substrate while we relate the increase in spin lifetime to improved contact properties according to our advanced transfer technique which has several advantages over the previous fabrication methods. At first, the contact region has not been exposed to an electron beam which may reduce the number of defects in graphene. \cite{teweldebrhan:013101, 10.1063/1.3502610} Secondly, the interface between MgO and graphene is expected to be of higher quality yielding more homogeneous barriers which can be seen by the larger $R_cA$ values for devices with thinner MgO layer thicknesses \cite{Droegeler2014}.

Our results indicate that the overall improvement of the spin properties primarily result from the improvement of the contact properties suggesting that the observed 1/$\mu$ dependence in the initial work is of extrinsic origin. In this context, we again want to emphasize (sections \ref{lifetime} and \ref{Mobilities}) that the determination of both the charge carrier mobility and the spin lifetime is by no means unambiguous and may thus result in contradicting $\tau_s - \tau_p$ dependencies.

Furthermore, we demonstrated in \cite{PhysRevB.88.161405} and \cite{PhysRevB.90.165403} that the spin lifetimes in \cite{PhysRevLett.107.047206} are most likely limited by contact properties (also see figure \ref{figure5}) in all devices. In fact, there are indications that even in our newest bottom-up devices the contacts may still limit spin transport properties (paper in preparation). If in current devices extrinsic parameters limit spin transport properties, it will be interesting to see how far more advanced fabrication methods will yield devices with even longer spin lifetimes and larger carrier mobilities which ultimately allows to unveil intrinsic spin scattering mechanisms.

\section{Acknowledgments}
The research leading to these results has received funding from the DFG through FOR-912 and the European Union Seventh Framework Programme under grant agreement no. 604391 Graphene Flagship.

\section*{References}

\bibliography{Literature}

\end{document}